\documentclass[reprint,prl,floatfix,nofootinbib,superscriptaddress]{revtex4-1}
\usepackage{color}
\usepackage[utf8]{inputenc}
\usepackage[T1]{fontenc}

\usepackage[colorlinks=true]{hyperref}% add hypertext capabilities
\usepackage{graphicx}% Include figure files
\usepackage{dcolumn}% Align table columns on decimal point
\usepackage{bm}% bold math
\usepackage{amsmath,amssymb,amsthm} 
\usepackage{amsfonts}
\usepackage{comment}
\usepackage{bbm}
\usepackage[usenames,dvipsnames]{pstricks}
\usepackage{subfigure}
\usepackage{braket}
\usepackage{float}
\usepackage{xcolor}
\usepackage{todonotes}
\DeclareFontFamily{U}{dutchcal}{\skewchar\font=45 }
\DeclareFontShape{U}{dutchcal}{m}{n}{<-> s*[1.0] dutchcal-r}{}
\DeclareFontShape{U}{dutchcal}{b}{n}{<-> s*[1.0] dutchcal-b}{}
\DeclareMathAlphabet{\mathlcal}{U}{dutchcal}{m}{n}
\SetMathAlphabet{\mathlcal}{bold}{U}{dutchcal}{b}{n}

\begin{document}

\title{Relative, local and global dimension in complex networks}

\author{Robert Peach}
\affiliation{These authors contributed equally.}
\affiliation{Department of Mathematics, Imperial College, London SW7 2AZ, UK}
\affiliation{Department of Neurology, University Hospital Würzburg, Würzburg, Germany}

\author{Alexis Arnaudon}
\affiliation{These authors contributed equally.}
\affiliation{Department of Mathematics, Imperial College, London SW7 2AZ, UK}
\affiliation{Blue Brain Project, École Polytechnique Fédérale de Lausanne (EPFL), Campus Biotech, 1202 Geneva, Switzerland}

\author{Mauricio Barahona}
\affiliation{Department of Mathematics, Imperial College, London SW7 2AZ, UK}
\affiliation{m.barahona@imperial.ac.uk}

\date{\today}

\begin{abstract}
Dimension is a fundamental property of objects and the space in which they are embedded. 
Yet ideal notions of dimension, as in Euclidean spaces, 
do not always translate to 
physical spaces, which can be constrained by boundaries and distorted by inhomogeneities, or to intrinsically discrete systems such as networks. 
To take into account locality, finiteness and discreteness, dynamical processes can be used to probe the space geometry and define its dimension.
Here we show that each point in space can be assigned a relative dimension with respect to the source of a diffusive process,
a concept that provides a scale-dependent definition for local and global dimension also applicable to networks. To showcase its application to physical systems, we demonstrate that the local dimension of structural protein graphs correlates with structural flexibility, and the relative dimension with respect to the active site uncovers regions involved in allosteric communication. 
In simple models of epidemics on networks, the relative dimension is predictive of the spreading capability of nodes, and identifies scales at which the graph structure is predictive of infectivity. We further apply our dimension measures to neuronal networks, economic trade, social networks, ocean flows, and to the comparison of random graphs.
\end{abstract}

\maketitle
\section*{Introduction}

One of the first forays into graph dimensionality originated with Erd\"os, when he explored the embedding of graphs into a minimum finite dimensional Euclidean space~\cite{erdos1965dimension}. This line of study helped realise the algorithmic importance of geometric interpretations of graphs~\cite{lovasz2019graphs} but was unfortunately no more than a by-product of the graph embedding process, yielding little actionable information~\cite{linial1995geometry}. Later, by characterising the fractal properties of complex networks, a measure of network dimension was defined in terms of the scaling property of a network topological volume~\cite{csanyi2004fractal,gastner2006spatial,shanker2007defining}.
Whilst the fractal approach showed that dimension plays an important role in characterising network topology and governing dynamical processes such as percolation~\cite{daqing2011dimension}, it was initially limited to global descriptions of network dimension. Extensions that considered the local scaling properties of the volume at different topological distances from a node were introduced in~\cite{silva2012local} and have been used to define a node-centric dimension that can identify influential nodes~\cite{pu2014identifying,bian2018identifying} or vital spreaders in infection models~\cite{wen2020vital}. 

However, methodologies based on fractal approaches assume that the topological volume follows a power-law distribution, a strong assumption, not necessarily accurate in real world networks exhibiting heterogeneities~\cite{gastner2006spatial}. Similarly, in classic papers such as~\cite{reuveni2010anomalies}, where the dimension of a node is defined using the decay rate of diffusion, or in~\cite{lacasa2013correlation}, where a random walk is used to create node embeddings, the same assumptions of homogeneity are required and an intermediate scale of dynamics must be chosen. As an an example, with a diffusive source located at the joining of a 1-d and a 2-d space, by measuring the decay rate we immediately ignore the heterogeneity of the space and simply find a dimension somewhere between $1$ and $2$. In this paper, we posit that the dimension at a node can, and should be, defined as \emph{relative} to another node. Using the solution of diffusion at other nodes relative to the source we are able to define a \emph{relative dimension}.

\section*{Results}
{\bf Graph dimension from diffusion dynamics.}
We start with the Green's function of the diffusion equation in $d$ dimensions
\begin{align}
    G_{t}(\mathbf x) = \left(4\pi \sigma t\right)^{-d/2} \exp\left ( - \frac{\|\mathbf x\|^2}{4 \sigma t}\right )\, ,
    \label{green}
\end{align}
which, together with an initial condition as a delta function at some position $\mathbf{x_0}$, provides a solution of diffusion equation as $p(\mathbf x, t) = G_t(\mathbf x - \mathbf x_0)$. From hereon, we refer to the time evolution of $p(\mathbf x, t)$ as the transient response.
As already considered in our previous works~\cite{peach2020semi,MSC}, these solutions have a maxima in their transient response at any other location $\mathbf{x}$, at time $\widehat t$ and amplitude $\widehat p$ given as
\begin{align}
    \widehat t( {\mathbf x}) = \frac{\|{ \mathbf x}\|^2}{2 d\sigma}\, , \qquad\widehat p(\widehat t) = (4 e \pi \sigma \widehat t)^{-\frac{d}{2}}\, ,
    \label{p_peak}
\end{align}
where, without loss of generality, $\mathbf{x_0}=0$. Then, the dimension at any point $\mathbf x$ relative to $\mathbf x_0$ can be evaluated to yield the definition of the relative dimension 
\begin{align}
    d(\mathbf x| \mathbf x_0) = \frac{ -2 \ln \widehat p}{\ln\left ( 4 e \pi \sigma \widehat t\,\right)}\, . 
    \label{relative_dimension}
\end{align}
Clearly, on the Euclidean space $\mathbb R^d$, the relative dimension is always equal to $d$, independently of $\mathbf x$ and $\mathbf x_0$.
However,
if we instead consider a compact subspace $\Omega \subset \mathbb R^d$, the diffusion dynamics will deviate from those prescribed in Equation~\eqref{green} due to the presence of boundaries relative to $\mathbf x$ and $\mathbf x_0$. 

The key property of Equation~\eqref{relative_dimension} that allows us to generalise it to graphs is that the positions $\mathbf x_0$ and ${\mathbf x}$ are not explicit in the right hand side but only used as labels to initialise the diffusion dynamics and measure the transient response.
Consequently, the relative dimension can be seen as intrinsic as it does not rely on any Euclidean embedding, but only on the existence of a diffusion dynamics on the original space.
In particular, on graphs we can use the standard diffusion process 
\begin{align}
    \partial_t \mathbf p(t) = -L \mathbf p\, , 
    \label{diff-network}
\end{align}
for a time-dependent node vector $\mathbf p(t)$ with $L$ the normalised graph Laplacian $L = K^{-1}(K - A)$ (corresponding to Euclidean diffusion in the continuous limit~\cite{singer2006graph}), where $K$ is the diagonal matrix of node degrees. Using a delta function at node $i$ with mass $m_i$, $\mathbf{p}(0) = (0, 0, \dots, m_i, \dots, 0)$,  as our initial condition, the $j$-th coordinate of the solution of Equation~\eqref{diff-network} (the so-called transient response of $j$) is given by the heat kernel
\begin{align}
    p_j(t|i) = m_i \left (e^{-tL}\right)_{ij}\, .
    \label{eq:network_solution}
\end{align}
By numerically solving~\eqref{eq:network_solution}, we can measure the time $\widehat t_{ij}$ and amplitude $\widehat p_{ij}$ at which a maximum appears in the transient response peak (time evolution) of node $j$ given a delta function initial condition at node $i$. 
In analogy to Equation~\eqref{relative_dimension},
we can then compute the full $N \times N$ matrix of relative dimensions with elements
\begin{align}
    d_{ij} = \frac{ -2 \ln \widehat p_{ij}}{\ln\left ( 4 e \pi \sigma \widehat t_{ij}\right)}\, .
    \label{network_relative_dimension}
\end{align}

\begin{figure*}[htpb]
    \centering
    \includegraphics[width=\textwidth]{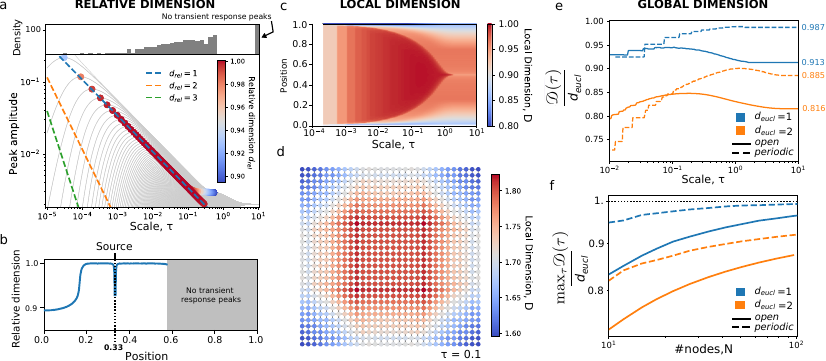}
    \caption{
    The relative, local and global dimension.
    {\bf a-c} Line graph example with $n=500$ nodes representing the interval $[0, 1]$.
    {\bf a} The relative dimension of nodes given a source located at $x=0.33$. The grey lines are the transient responses of the (non-source) nodes and the position of the peaks in the transient responses are highlighted by dots, coloured by their relative dimension.
    Top inset, a histogram of transient response peaks where the far right bin corresponds to nodes where no peak in the transient response was observed, and thus no relative dimension could be calculated.
    {\bf b} The relative dimension as a function of position in $[0, 1]$ shows a plateau near $d_{rel}=1$ for nodes near the source. The grey region indicates the set of nodes for which no peak was observed. 
    {\bf c} The local dimension of each node as a function of scale, where above $\tau =1$, the stationary state is attained and the local dimension is stable.
   {\bf d} The local dimension of the grid graph ($n=500$) at scale $\tau=0.1$, showing inhomogeneities due to the boundaries similar to the line graph.
   {\bf e} The evolution of the global dimension $\mathlcal D(\tau)$ (normalised by the expected Euclidean dimension) as a function of scale for the same line and grid graph as well as their periodic equivalent graphs, illustrating differing behaviors emerging from the influence of the boundaries or the topology.
   {\bf f} For the same graphs as in {\bf e}, we increase the number of nodes in each dimension to measure the convergence rate of 
   $\mathrm{max}_\tau \mathlcal D(\tau)$ to the underlying Euclidean dimension $d_{eucl}$, showing a faster convergence for lower dimensional spaces and periodic grids.
   }
    \label{fig:geometric}
\end{figure*}

To illustrate the notion of relative dimension, we used a line graph (Figure~\ref{fig:geometric}a-b) as a discrete representation of the continuous 1-D interval.
We observe that due to the boundaries, a large fraction of nodes do not have a peak in transient response, however for nodes near the source, where the boundary has no influence, the relative dimension is close to the expected $d=1$. We emphasise that the dimension is not derived from a fit to the data, as is common in measures of fractal dimensions~\cite{csanyi2004fractal,gastner2006spatial,shanker2007defining}, but instead is directly observed at the transient response relative to a source node.

It is then natural to define the \emph{local dimension} of a node $i$ by averaging the relative dimension of the nodes displaying a peak in their transient responses relative to $i$ before a given time $\tau$ as
\begin{align}
    \mathcal D_i(\tau) = \frac{\sum_{j=1, j\neq i}^n d_{ij}(\tau)\mathbbm 1_{\widehat t_{ij} < \tau}}{\sum_{j=1, j\neq i}^n\mathbbm 1_{\widehat t_{ij} < \tau}}\, , 
\end{align}
where $\mathbbm1_{\widehat t_{ij} < \tau}$ is the indicator function.
Whilst the local dimension can be likened to a measure of centrality, it also directly captures the dimension of the local embedding space.
In Fig.~\ref{fig:geometric}c we observe the increasing effect of the boundaries on local dimension as we increase the scale. 
Near the center of the line, and when considering nearby nodes (at short scales), one can expect to estimate a dimension near $1$, or equivalently $2$ for the grid shown in Fig.~\ref{fig:geometric}d. We observe in Fig.~\ref{fig:geometric}c a central region with $\mathcal D_i \sim 1$ that becomes increasingly smaller as scale $\tau$ increases; at short scales, the central region is insensitive to the boundaries since the diffusion has not yet reached them.
This `boundary insensitive central region' collapses at $\tau=1$ (corresponding to the spectral gap of the graph) when all nodes have aggregated information about the boundaries of the line graph. 

Finally, we can define a graph measure of dimension by averaging the local dimensions across multiple scales to obtain the global dimension
\begin{align}
    \mathlcal{D}(\tau) = \frac{1}{n} \sum_{i=1}^n \mathcal{D}_i(\tau) \, , 
    \label{global-d}
\end{align}
still dependent on $\tau$.
In Fig.~\ref{fig:geometric}e we display the global dimension (as a ratio to the expected Euclidean dimension) for the line and grid graphs and their periodic equivalents (the circle and sphere graphs respectively). 

Whilst the periodic equivalents do not contain boundaries, they are still constrained to a compact space that will introduce topological effects, e.g., on a periodic graph the diffusion will interact with itself at the opposite side to the initial condition. We first notice that the non-periodic graphs display a maximum in global dimension, likely when the effect of the boundaries is lowest.
In contrast, the periodic graphs do not exhibit a peak of the same magnitude suggesting that the topological effect of a compact space has less impact on the global dimension than the presence of a boundary.

In the context of graphs as discrete Euclidean spaces, the maximum of the global dimension curve (Fig.~\ref{fig:geometric}e) can be seen as an approximation of the Euclidean dimension, whereas the global dimension at largest scale characterises the effect of the boundary or topology of the graph. It should be noted that for a non grid-like graph, what is a boundary or a topological effect is not clear.
By increasing the graph size, and thus reducing the effects of the boundaries, the global dimension converges towards the expected Euclidean dimension (Fig.~\ref{fig:geometric}f). For the grid, the surface of the boundary increases with respect to the volume of the space and results in a slower convergence, whereas the global dimension of the periodic grid is only affected by the topology, and thus converges faster.

{\bf Delaunay meshes and inhomogeneities.}

\begin{figure*}[htpb]
    \centering
    \includegraphics[width=\textwidth]{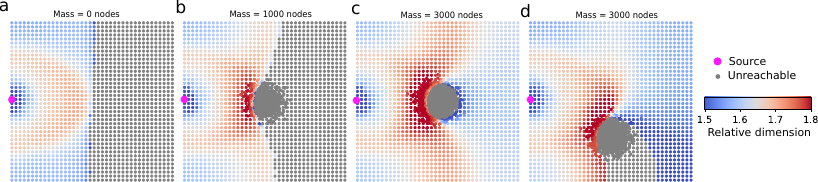}
    \caption{
    Inhomogeneities and lensing.
    The relative dimension from a point source (pink dot) to other nodes in a Delaunay grid graph (edges not shown). 
    Grey nodes indicate nodes for which no transient response peak was detected. The regular grid is shown in {\bf a}, and an additional mass is added in {\bf b-d}, with varying mass and position, showing a effect similar to gravitational lensing.
   }
    \label{fig:lensing}
\end{figure*}

To develop more intuition for our measure of relative dimension, we consider a simple constructive example using Delaunay meshes in Figure~\ref{fig:lensing}. Given a source-node located at the left boundary of a homogeneous delaunay mesh, relative dimension displays an inhomogeneous distribution radially from the source until nodes do not have a transient response peak (Figure~\ref{fig:lensing}(i)). 
Adding nodes near the centre of the Delaunay grid graph creates local inhomogeneities modifying the underlying space, with a clear analogy to the theory of gravitation and gravitational lensing~\cite{einstein1936lens}. In particular, the added mass acts as a gravitational lens for the diffusion process, whereby nodes directly behind the point mass that were previously 'unreachable' can be 'reached by the diffusion' if the mass is sufficiently large. Small masses are reminiscent of weak lensing (Figure \ref{fig:lensing}(ii)), whereas larger masses are closer to strong lensing (Figure \ref{fig:lensing}(iii))~\cite{misner1973gravitation}. The behaviour of relative dimension in the presence of inhomogeneities suggests that diffusion effectively occurs on a curved geometry induced by the presence of the mass. Moving the mass towards one boundary (Figure \ref{fig:lensing}(iv)) shows some coupling between the lensing effect and the presence of the boundary. 
All three possible effects, boundaries, topology and inhomogeneities, are thus important in the notion of dimensions, but may not be distinguishable in more complex networks.
Nevertheless, our notion of relative dimension is able to capture them all in one graph-theoretical measure.

{\bf Dimensions in protein structure: rigidity and allostery.}

\begin{figure}[htpb]
    \centering
    \includegraphics[width=\columnwidth]{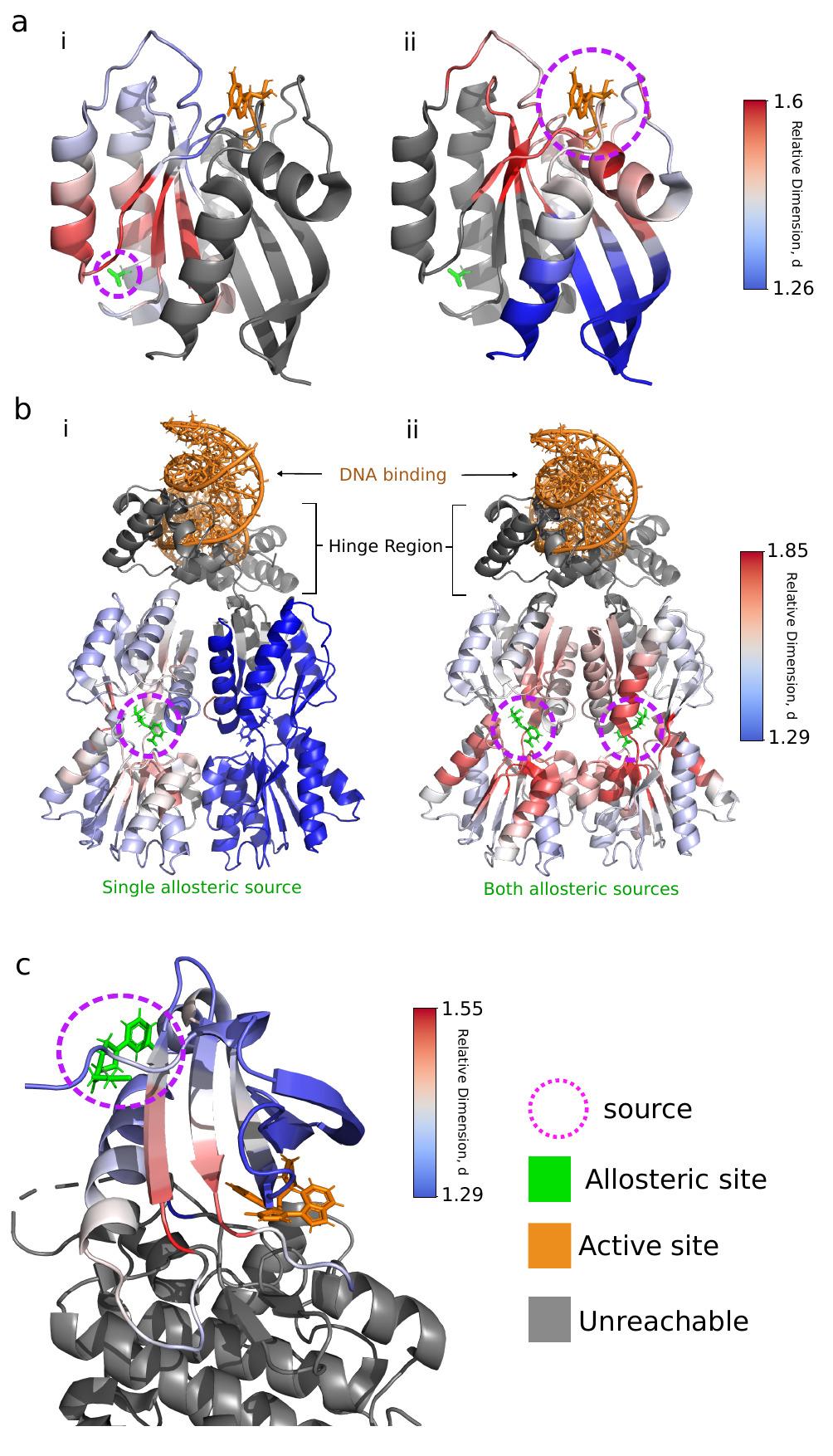}
    \caption{
    Relative dimensions in allosteric proteins. Protein residues (amino acids) are coloured according to their relative dimension to the source region, where grey indicates that no peak in the transient response was identified and thus relative dimension could not be calculated.
    \textbf{a} The relative dimensions of all atoms in HRas GTPase (PDB ID: 3K8Y) given (i) the allosteric site and (ii) the active site as the source of diffusion.
    \textbf{b} Relative dimension in the multi-allosteric site Lac Repressor protein (PDB ID: 1EFA) given a (i) single allosteric site source and (ii) for both allosteric sites simultaneously.
    \textbf{c} The relative dimension give the allosteric site as the source in PDK1 (PDB ID: 3ORX).
    }
    \label{fig:figure_allostery}
\end{figure}

We then apply the relative dimensions on a real-world example with allostery in proteins, a phenomena whereby a subset of a protein (active site) can be modulated (activated or inactivated) through binding of a ligand at another subset of the protein (allosteric site). 
We examine three well-studied allosteric proteins: HRas GTPas, Lac repressor and PDK1
in Figure~\ref{fig:figure_allostery} (for more details on these proteins, see Methods). In HRas, we find a low relative dimension at the active site given the allosteric site as the source (Fig.~\ref{fig:figure_allostery}a(i)), but in reverse the allosteric site does not see a transient peak from the diffusion started in the active site (Fig.~\ref{fig:figure_allostery}a(ii)). 
Even if an exact statement of allosteric mechanism is not our purpose here, it is interesting to note that a low relative dimension suggests a more `direct' or `funneled' communication from the allosteric site to the active site. Moreover, the asymmetry of this communication may relate to different functions for each half of the protein.

The lac repressor protein is constructed from two separate monomers and it is generally understood that binding of both NPF molecules (one on each monomer) is required to activate the lac repressor via a cooperative allosteric effect acting on the hinge region~\cite{muller1996lac}.
Given that the allosteric mechanism is cooperative, we do not expect a direct communication to the active site from the allosteric site, and instead we examined the change in relative dimension upon using a single allosteric site as a source (Fig.~\ref{fig:figure_allostery}b(i)) vs. both allosteric sites as sources simultaneously (Fig.~\ref{fig:figure_allostery}b(ii)).
We find that when binding NFP to just one monomer the relative dimension across the entire protein is lower when compared to using both allosteric sites as sources of diffusion. 

Finally, binding at the PDK1 interacting fragment (PIF) on PDK1 triggers a signal to start the phosphorylation of the activation loop of the substrates at the ATP pocket, or active site~\cite{biondi2001pif}, and thus we would expect direct communication between the active and allosteric sites.
Using the allosteric site as the source of our diffusion (Figure~\ref{fig:figure_allostery}c), we find that a large region of PDK1 does not return a relative dimension (grey region in Figure~\ref{fig:figure_allostery}c). We remind the reader that to calculate relative dimension we must observe a peak in the transient response. Of those residues for which relative dimension was computed, the activation loop displays the lowest relative dimension to the allosteric site. We hypothesise that a lower dimension pathway from the allosteric to active site will improve the efficiency of communication transfer since it becomes more direct.

\begin{figure*}[htpb]
    \centering
    \includegraphics[width=2\columnwidth]{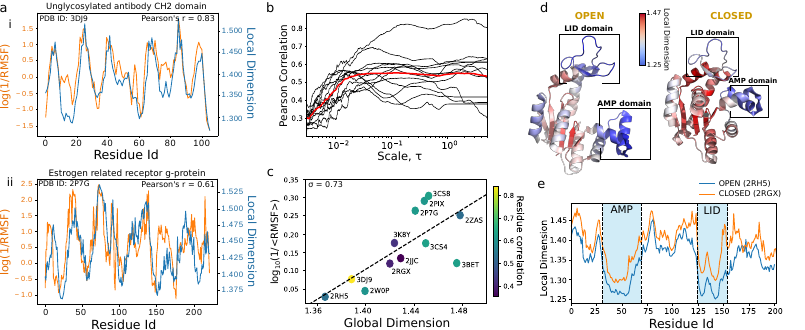}
    \caption{
    The relationship between root-mean-square fluctuations (RMSF) of protein residues and their local and global dimension. 
    {\bf a} The log-inverse RMSF vs local dimensionality for each residue in 
    {\bf i} unglycosylated antibody CH2 domain, 
    {\bf ii} Estrogen Related Receptor g.
    {\bf b} The positive correlation ($0.539$) between local dimension and log-inverse RMSF at the residue level across $12$ different proteins as a function of scale.
    {\bf c} A strong positive correlation between global dimensionality of $12$ proteins against log-inverse RMSF. 
    {\bf d} The local dimension of each residue (i) mapped onto Aquifex Adenylate Kinase in the open (PDB ID: 2RH5) and closed (PDB ID: 2RGX) conformations and (ii) plotted by residue id.
    }
    \label{fig:figure_rmsf}
\end{figure*}

Whilst the relative dimension provides insights into allostery, we can leverage the local and global dimension to examine protein dynamics. 
In Figure~\ref{fig:figure_rmsf}(a), we show a strong correlation between the local dimension and $\log_{10}(1/\text{RMSF})$ of residues for Figure~\ref{fig:figure_rmsf}a(i) an unglycosylated antibody CH2 domain and Figure~\ref{fig:figure_rmsf}a(ii) an Estrogen Related Receptor g protein. The results here suggest that a residue with a larger local dimension is associated with a lower flexibility and thus lower degrees of freedom. 

To examine this further, we plotted the Pearson correlation between local dimension and $\log_{10}(1/\text{RMSF})$ for $12$ randomly chosen proteins in Figure~\ref{fig:figure_rmsf}(b).
We see that at middling to long timescales of diffusion the correlation plateau with an average at about $\sigma=0.55$ suggesting that the relationship between local dimension and protein flexibility is robust.
Calculating the global dimension for the same set of proteins in Figure~\ref{fig:figure_rmsf}(c), we find a correlation (Pearson $\sigma=0.73$) between global dimension and the $\log_{10}(1/\langle\text{RMSF}\rangle)$ of a protein. The global values of dimension sit between $1.36$ and $1.5$ for the $12$ proteins.
These results agree with studies that show spectral dimension is generally $<2$ and decreases with an increase in flexibility~\cite{reuveni2010anomalies,reuveni2008proteins}.

We now take a deeper look at \emph{Aquifex} Adenylate Kinase (ADK), a dynamical protein with three subdomains: the lid, AMP and core domains.
We find that the closed conformation displays a higher local dimension due to the presence of stabilising interactions, not present in the open conformation, creating a more compact structure (Figure~\ref{fig:figure_rmsf}d). The AMP and lid domains are known to open and close around substrate. We find that both have a lower local dimension relative to the core domain (Figure~\ref{fig:figure_rmsf}e) and that the AMP domain to have a lower average local dimension than the lid domain in both conformations. The latter we validated using experimental fluorescence correlation spectroscopy that shows that the AMP domain to open and close at a faster rate ($16.2 \mu s$) than the lid domain ($46.6 \mu s$)~\cite{peach2017exploring,peach2019unsupervised}.

\begin{figure*}[htbp]
    \centering
    \includegraphics[width= 0.9\textwidth]{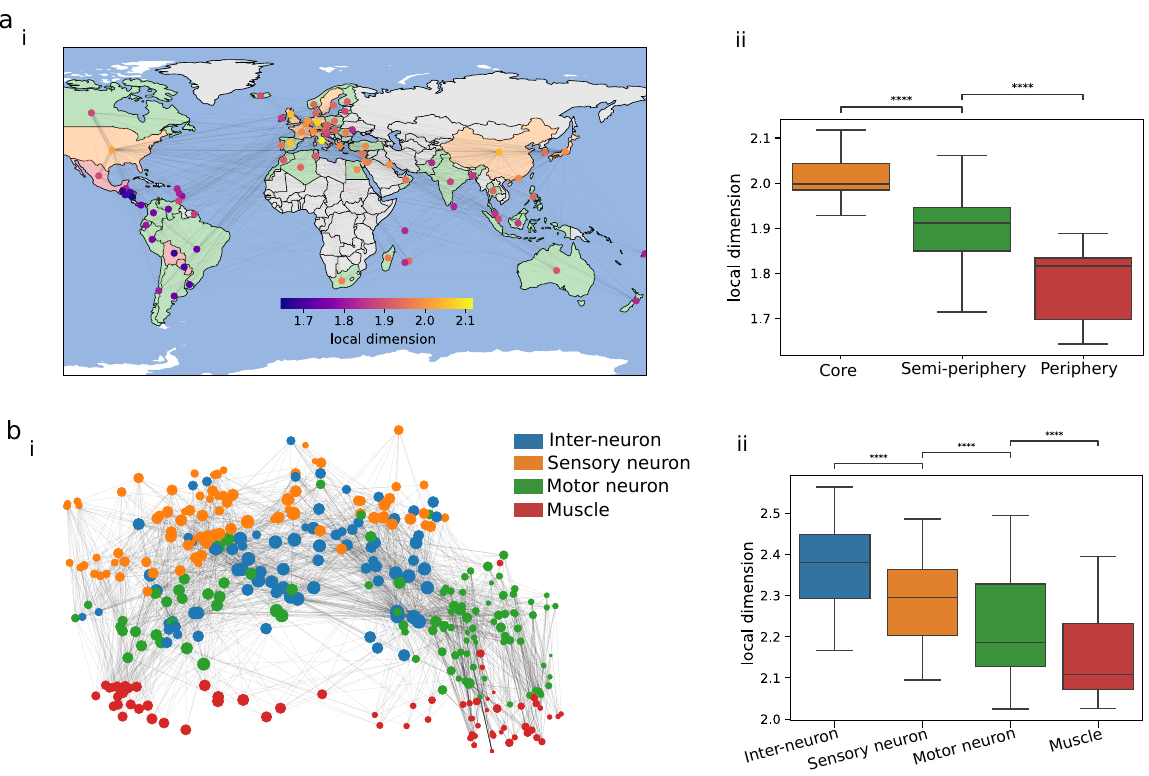}
    \caption{Spatially embedded networks with long range interactions.
    {\bf a} {\bf i} The local dimension (coloured nodes) of the 1994 world trade data set ~\cite{de2018exploratory} ($\tau=0.57$). Countries are coloured according to their position in the world partition (grey countries not in data set).
    {\bf ii} Comparison of local dimension by world partition shows a decreasing local dimension as the countries become more peripheral.
    {\bf b} The \textit{C.\ elegans} connectome consisting of $279$ neurons and $98$ muscles coloured by neuron type/muscle are shown in {\bf i} . 
    In {\bf ii}, the boxplots of local dimensions at large scale show that the neuron types statistically differ.
    All statistical tests were Mann-Whitney-Wilcoxon test two-sided with Bonferroni correction, $****$: p $\leq 0.0001$).
    }
     \label{fig:figure_complexnetworks}
\end{figure*}

{\bf Local dimension as a means to differentiate node roles.}
To further explore our measure of dimension in the context of identifying roles of nodes within the network, we present two examples of real-world complex networks in Figure~\ref{fig:figure_complexnetworks} where nodes have pre-assigned roles. 
The first example explores the world trade network (consisting of $80$ nodes) of metal manufacturing in $1994$~\cite{de2018exploratory}, where nodes correspond to countries and directed incoming edges represent the amount of weighted imports from another country. A well established concept in economic theory partitions countries based on their positioning (1. core, 2. semi-peripheral, 3. peripheral) within the world economy~\cite{smith1992structure}. 
For the largest scale, we find significant differences between distributions of the local dimension for each of the world partitions (Fig~\ref{fig:figure_complexnetworks}b). 
There is almost no overlap in local dimension between the two extreme partitions, core and periphery, but the distribution of local dimension for semi-peripherical nodes is wider, suggesting that this class of countries is more diverse.

Our second example is the undirected connectome (N=377) of the nematode \textit{C.\ elegans} (Figure~\ref{fig:figure_complexnetworks}b(i)) with the inclusion of muscles, important for examining control~\cite{yan2017network}(\url{https://www.wormatlas.org/neuronalwiring.html}), and where scales have previously been shown as important~\cite{bacik2016flow}.
We compare the dimension of the three different neuronal types (inter neurons, sensory neurons, motor neurons) and muscles, at long scales in Figure~\ref{fig:figure_complexnetworks}b(ii), and find significant differences in their local dimensions. Inter-neurons, are central nodes of neural circuits that enable communication between sensory and motor neurons, thus we would expect them to sit in a higher-dimensional space, whereare muscles are peripheral as they display the lowest local dimension, likely aiding with the direct propagation of signals. In addition, we find the the highest dimensional nodes are the important control motor neurons AVA/AVB neurons (both left and right), resulting in uncontrolled motion if ablated~\cite{yan2017network} (see Supplementary Table 1 for top $40$ local dimension neurons). 

{\bf Local dimension as scale dependent measure of centrality.}
Measures of centrality are some of the most fundamental tools in network theory. Here, we show that the local dimension can also be utilized as a scale-dependent centrality measure, such as those derived in~\cite{MSC, estrada2008communicability}.
To illustrate the use of the local dimension as a centrality measure for complex networks we analyzed two datasets where the importance of nodes changes substantially with scale.

\begin{figure*}[htbp]
    \centering
    \includegraphics[width= 0.9\textwidth]{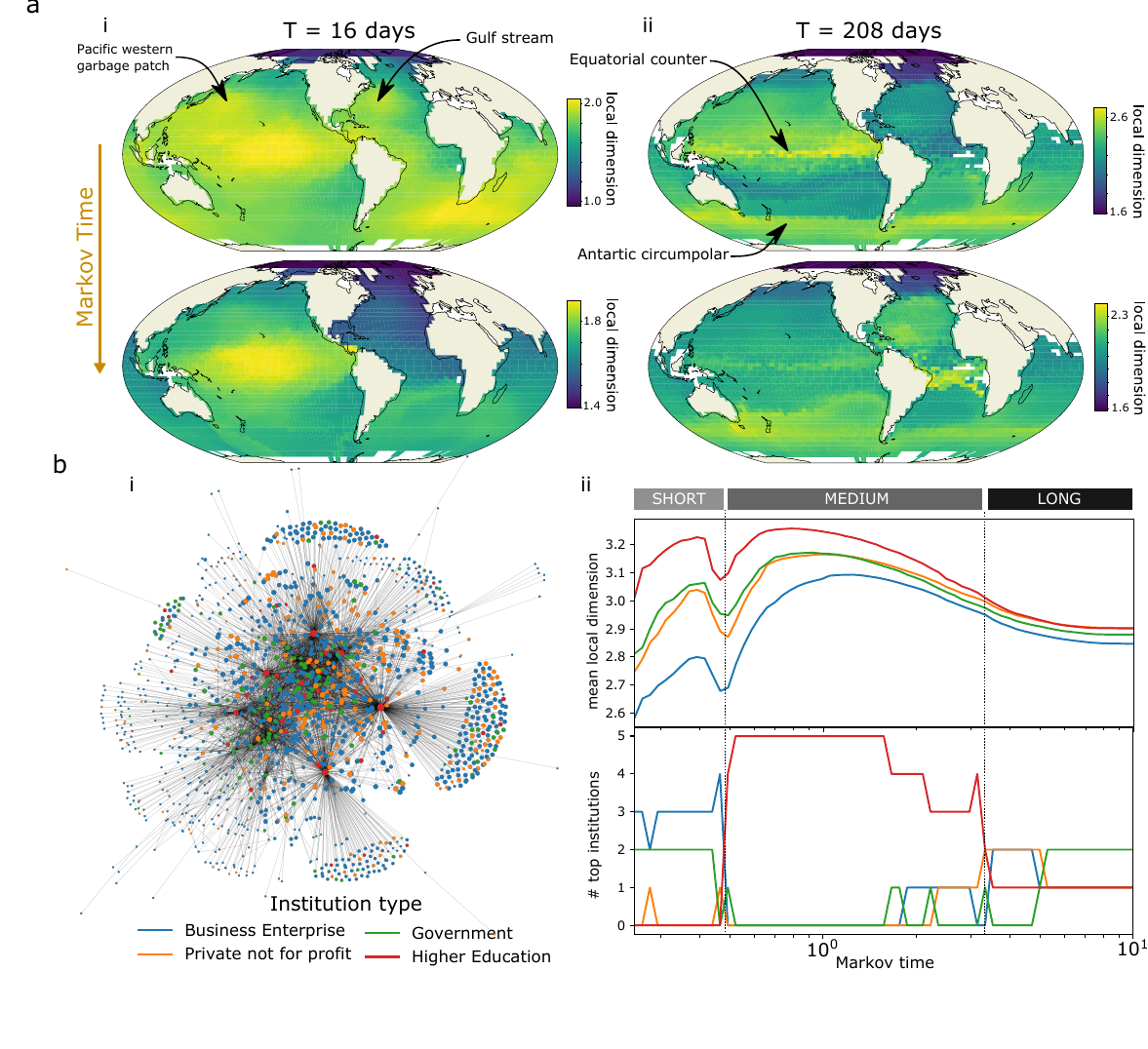}
    \caption{Illustration of local dimension at several scales in two dataset. In {\bf a}, we considered two graphs extracted from the ocean drifter by~\cite{faccin2021state} where edges are the number of drifters crossing two regions of the ocean within {\bf i} $T=16$ and {\bf ii} $208$ days. For each, we selected a small and a large scale of local dimension, each representing various known features of the ocean dynamics, mostly located between the main gyres such as between the north and south equatorial in the pacific, or along the antarctic circumpolar current.
    In {\bf b} we explored a {\bf i} social network of scientific collaborations between New Zealand institutions {\bf ii} across scales, to find that the top $5$ institutions are businesses at small scales, universities at longer scales, and a mix of institutions at stationarity.
    From middle scales, the University of Auckland remains the top ranked node until stationarity.
    }
     \label{fig:figure_centrality}
\end{figure*}

First we look at the global network of ocean surface currents derived from the Global Drifters program (\url{http://www.aoml.noaa.gov/phod/gdp/index.php}) constructed by~\cite{faccin2021state}(\url{https://github.com/maurofaccin/ocean_surface_dataset}). Each node is associated with a small region of the ocean, and an edge between two nodes counts the number of drifters passing from one to another region in a given time interval $T$. For short times, such as $T=16$ days, the graph connectivity remains local with respect to the spatial embedding of the nodes on the earth surface, but with larger times ($T=208$ days) the connectivity becomes long range and complex (see also the degree distribution in Supplementary Figure 1).
We can examine both time intervals at short and long scales of our local dimension (Figure~\ref{fig:figure_centrality}a); the small or large scale local dimension provide different perspectives on regions of high dimensions, related to regions where the ocean flow has a more complex dynamics. At small time intervals and short scales (Figure~\ref{fig:figure_centrality}a(i) top), we identify locally high dimension regions such as the Gulf stream or the Pacific garbage patch where drifters remain trapped and circulate quickly. If we look at long time intervals (Figure~\ref{fig:figure_centrality}a(ii)), we notice bands of high dimension which represent the boundaries between main gyres, such as that along the equator. At short scales, the drifters have lower dimensional dynamics while they follow these currents. However, at longer scales the drifters can drift north or south of the equator and be further transported to widely different regions throughout the world, and thus the dimension of the boundaries between major ocean currents is larger. We also note a visual similarity between the small time interval and long scale (Figure~\ref{fig:figure_centrality}a(i) bottom) and long time interval and short scale (Figure~\ref{fig:figure_centrality}a(ii) top), whereby the drifter movements are generally split between north and south.
Our results provide further evidence that a notion of scale in the analysis of ocean flow is crucial to exploit and interpret the dynamics~\cite{faccin2021state}.

Finally, we examine a complex social network of scientific collaborations between New Zealand institutions (Figure~\ref{fig:figure_centrality}b). Each node represents an institution which falls into the following categories: Higher education, Government, Private not for profit, or Business Enterprise.
Edges are weighted by the number of collaborations between two institutions in the time period 2010-2015, measured by co-authored publications on Scopus~\cite{aref2018analysing}.
We compute the local dimension as a function of scale on this network and identify three main scales (short, medium and long; Figure~\ref{fig:figure_centrality}b(ii)).
On average and across scales, the higher education institutions displayed the highest local dimension and business enterprises were lowest. However, if we instead look only at the $5$ nodes with the highest local dimension, we find that at short scales, businesses and government institutions comprised the top $5$ local dimension nodes, highlighting their high dimension to a small neighbourhood. For a wide range of medium timescales, we find that the universities display the largest local dimension, reflecting their hub like role in the network (Figure~\ref{fig:figure_centrality}b(i)). 
At long timescales (in the limit close to stationarity) we find a mixture of nodes from all institutions appear in the top $5$ nodes.
A previous study used betweenness and eigenvector centrality to show that most central institutions were not solely universities, but was also comprised of other institution types~\cite{aref2018analysing}.
Here, we show that the precise role of each node depends on the choice of scale, as already discussed in ~\cite{MSC}.

\begin{figure*}[htbp]
    \centering
    \includegraphics[width= \textwidth]{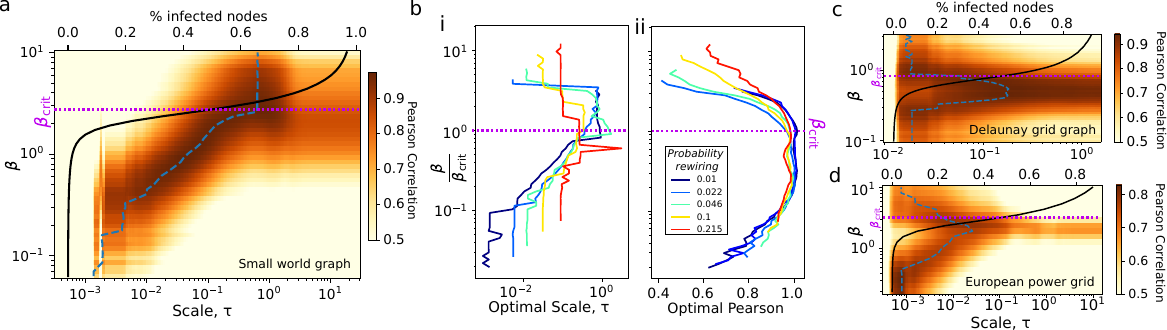}
    \caption{
    Dimension and epidemic spreading.
    \textbf{a} Heatmap of Pearson correlation between local dimension and node infectiousness for small-world graph ($n=100$, average degree $k=10$, probability of rewiring $p=0.015$). The black line is the average proportion of infected nodes given a single-seed node for a given infection probability $\beta$. The transition from low to high proportion of infected nodes indicates the critical point while the dashed line is the maximum correlation for each $\beta$.
    \textbf{b} We vary the probability of rewiring edges $p$ of small world graphs and display {\bf i} the diffusion time that maximises the correlation between local dimension and infectiousness for varying $\beta$, and {\bf ii} the associated correlation coefficient. The correlation is near one close to criticality and above $0.8$ for a large range of $\beta$.
    We repeat the analysis in \textbf{a} for \textbf{c} a Delaunay grid graph ($n=400$) and \textbf{d} the European powergrid network to observe a similar linear relationships between scale and infection probability $\beta$ prior to criticality.
    }
     \label{fig:figure_sir}
\end{figure*}

{\bf Dimension in epidemic spreading.}
What about dynamical processes on networks? In Fig.~\ref{fig:figure_sir}a we use an SIR model on Watts-Strogatz small-world networks~\cite{watts1998collective} and by scanning the infection probability $\beta$, we show that the local dimension of a node strongly predicts its infectiousness.
Below the critical regime of large infectiousness, we find that infection probability is positively correlated with the scale, i.e. the size of the local neighbourhood that should be considered grows with the infection probability. 
However, near criticality $\beta_\mathrm{crit}$ (a threshold infection probability), we observe a behavior similar to a phase transition, whereby the time scale that local dimension correlates best with node infectiousness diverges towards values near unity, corresponding to the largest scale of the local dimension. 

We further computed the local dimension and SIR dynamics for small-world graphs whilst varying the probability of rewiring $p$ parameter, to interpolate between near regular graphs to Erd\H{o}s-R\'eyni random graphs. In Fig.~\ref{fig:figure_sir}b we observe that the relationship between the optimal scale to determine local dimension and infectiousness of a node disappears with the randomness of the network.
At low $\beta$, node infectiousness is dominated by the distance from high degree nodes in a small-world graph and, as $\beta$ increases, the spreading dynamics accelerates and nodes further away can be infected. A local dimension at longer time scales $\tau$ is therefore necessary to obtain a better prediction on node infectiousness.
However, in Erd\H{o}s-R\'eyni random networks all nodes are on average at equal distance from high degree nodes and no meaningful scale exists.

We find similar linear relationships between $\beta$ and scale in a Delaunay grid graph (Fig.~\ref{fig:figure_sir}c) and the European power grid (Fig.~\ref{fig:figure_sir}d). 
The decrease in scale for the local dimension to be a good predictor beyond $\beta_\mathrm{crit}$ for both graphs echoed the results of high probability re-wiring in small-world graphs, suggesting that global graph structure becomes less important if the infection probability is sufficiently high.

{\bf Graph classification from distributions of local dimensions.}

\begin{figure*}[htbp]
    \centering
    \includegraphics[width= 0.8\textwidth]{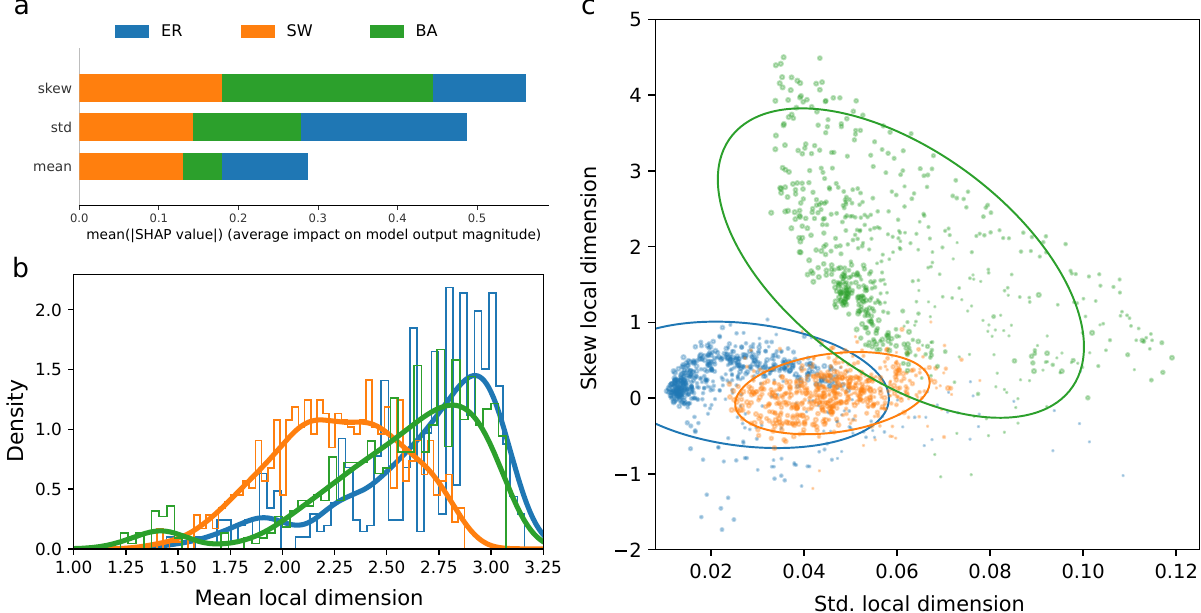}
    \caption{
    Comparison of random networks (Erdos Reyni, Watts-Strogatz, Barabasi-Albert) by features derived from the distribution of local dimension (mean, standard deviation, skewness). For each graph type, $600$ graphs with varying choices of parameters was chosen (see Methods).
    {\bf a} The shap value for each feature reveals the importance of each feature in distinguishing random graph types in the trained random forest model.
    {\bf b} A density histogram revealing the distribution of the mean local dimension of each random graph type.
    {\bf c} Each graph is plotted by their skewness and standard deviation of local dimension, coloured by the random graph type and marker size is proportional to the number of nodes. The ellipses indicate a $2$ standard deviation confidence interval using a Pearson correlation for each graph type.
    }
     \label{fig:figure_graphclassification}
\end{figure*}

Random graphs, such as the Watts-Strogatz graph used above, sit at the intersection of graph theory and probability theory, and are often used to investigate the properties of `typical' graphs. 
Various models of random graphs exist to cover the diversity of complex networks encountered in the real-world, but the most commonly discussed are Erd\H{o}s-R\'eyni, Watts-Strogatz, and Barabasi-Albert graphs. 
To understand whether the distribution of local dimension differed across these three types random graphs, we generated a large data set with various choices of parameters to generate each type of random graphs of similar sizes (see Methods).
We then computed the local dimension of each node of each graph and extracted three features from the distribution of local dimension (mean, standard deviation and skewness) and used a Random Forest model to classify between the random graph types. 
The classification model achieved $0.95\pm 0.014$ accuracy with a stratified $10$-fold split, suggesting that different random graphs types display inherently different dimensional properties. 
A Shap feature importance analysis revealed that the skewness and standard deviation of the distributions were most informative in differentiating the random graph types (Figure~\ref{fig:figure_graphclassification}(a)). 
The skewness and standard deviation of Barabasi-Albert graphs were larger reflecting their extremely broad and non-homogenous degree distribution. 
As expected, an overlap in the distribution of Erd\H{o}s-R\'eyni and Watts-Strogatz graphs is observed (Figure~\ref{fig:figure_graphclassification}(c)) owing to the fact that Watts-Strogatz graphs were designed specifically to interpolate between lattices and fully disordered states (similar to, but not exactly Erd\H{o}s-R\'eyni~\cite{maier2019generalization}) via a rewiring of edges. 
Despite their overlap, Erd\H{o}s-R\'eyni graphs display a smaller standard deviation, likely resulting from a more homogeneous degree distribution.

\section*{Discussion}
In this paper we have introduced a new framework to define notions of dimensions not only on graphs, but on any space where a dynamical process (from which the Euclidean dimension can be inferred) can be defined. 
Our measure of dimension is defined using consensus dynamics on graphs, which is most similar to Euclidean diffusion, and naturally links with the dimension in the d-dimension diffusion equation. In this sense, our measure is intrinsically defined through the diffusive process taking place on a discrete system and recovers the intuitive definition of dimension as the system loses its discreteness. In doing so, we are also able to give a geometric meaning (through the notion of dimension) to the effect of boundaries and density inhomogeneities.
We have shown the relevance of this approach to examine real world systems such as protein dynamics, neuronal or social networks, ocean currents or epidemic spreading by examining the underlying graph structure.

Through various detailed studies with the relative dimension, probing local dimensions at various scales, or characterising entire graphs with the global dimension, we have provided evidence for the wide applicability of our dimension measures to both non-complex and complex networks (see SI for characterisation of degree distributions of graphs used in this paper).
There are a variety of practical applications where probing network geometry is of great utility~\cite{boguna2021network} and are within the scope of these dimension measures.
For example, spatially modulated neurons (such as place cells or grid cells), whose network architecture plays a fundamental role in the representation of space and spatial memory, could be studied with our measures to understand the local and global lattice arrangement of firing fields~\cite{ginosar2021locally}. 
Alternatively, our measures could be used to provide insights into the manifestation of material properties. For example, the angle at which two stacked layers of graphene are oriented relative to each other dictates the presence of superconductivity and fragile topology~\cite{cao2018unconventional}.
Further analysis of graph classification problems using the distribution of dimension measures (relative or local) are also promising in view of our preliminary results using random generative networks.

\section{Methods}

{\bf Graph diffusion.}
A network (or a graph) $G$ is a tuple $G=(\mathcal{V},\mathcal{E})$, consisting of the set of nodes $N=|\mathcal{V}|$ vertices and $M = |\mathcal{E}|$ edges connecting them. The network can be described by its $N\times N$ adjacency matrix which indicates the existence and the weight of a connection (edge) between each pair of nodes. 
On a graph, there are several non-equivalent definitions of diffusion, which are defined by different forms of the graph Laplacian. However, only one forms corresponds to the Euclidean diffusion, described by the normalised Laplacian $L = K^{-1}(K - A)$ where $K$ is the diagonal matrix of weighted degrees and $A$ the weighted adjacency matrix~\cite{singer2006graph}.
Using the definition of the Laplacian, we can state the diffusion equation for a $N \times 1$ time-dependent node vector $\mathbf p(t)$ as in Equation.\eqref{diff-network}, which is also known as consensus dynamics~\cite{masuda2017random}. 
For an initial condition with a delta function of mass $m$ at node $i$, the $j$-th coordinate of the solution of Equation~\eqref{diff-network} is given by Equation.~\eqref{eq:network_solution}.
For comparability across different graphs, we normalise the times of diffusion by the second smallest eigenvalue of the graph Laplacian, $\lambda_2$ (the spectral gap), thus $\tau=1$ is the timescale for the diffusion to reach stationarity.

From  our choice of Laplacian, the relative dimension matrix $d$ (that we introduce in the next section) is symmetric if the initial masses $m$ are chosen inversely proportional to the weighted node degrees.

In addition, to ensure that the stationary state of the diffusion sums to unity, we take $m_i = \overline k / ( n k_i)$
where $\overline k$ is the mean weighted degree and $n$ is the number of nodes in the source.
This is used in the protein example, where the initial mass are distributed on all the atoms of the allosteric or active site.

{\bf Comparison with fractal dimension}
Looking more closely at our definition of relative dimension of Equation~\ref{network_relative_dimension}, it is proportional to the ratio of natural logarithms of peak amplitude and time, which displays similarities to the fractal based approaches where an approximate dimension can be derived from the ratio of natural logarithms of mass at a radius $r$,
\begin{align}
    d \sim \frac{\log(M)}{\log(r)} \, , 
    \label{fractal_dimension}
\end{align}
where the mass $M$ is simply the number of nodes within some link distance $r$~\cite{daqing2011dimension}.

{\bf Computational aspects.}
Python code to compute the relative, local and global dimensions is available at \url{https://github.com/barahona-research-group/DynGDim}, based on the package NetworkX and numpy/scipy standard libraries.

{\bf Delaunay mesh with mass.}
We apply Delaunay triangulation to a $40$ by $40$ grid to return a weighted planar graph for which no point is inside the circumcircle of any triangle. The size of the grid is one unit of the code distance units.
We define the weights of each edge as the inverse Euclidean lengths between points and thus obtain a discretisation of the plane.
To simulate the gravitational lensing effect, we added additional nodes sampled from a Gaussian distribution with parameters with variance $0.05$ in the unit square with various positions and number of nodes.

{\bf Protein Graph Construction.}
The graph representation of the proteins used in this work are computed using~\cite{bagpype}, an extension of \cite{amor2016prediction}.
In short, from a pdb file, each atom is represented by a node, and bonds between atoms by an edge weighted by the energy of the bond. The choice of bonds is key to create a meaningful graph representation, and is explained in~\cite{amor2016prediction,bagpype}, see \cite{proteinlens} to access the code.

{\bf Root-mean square fluctuation calculations.}
Enzymatic proteins are inherently flexible and known to exhibit motions across a wide range of temporal and spatial scales. Using simulations, each atom can be assigned a root-mean square fluctuation (RMSF). We calculate the RMSF using the CABS-flex 2.0 webserver which simulates protein dynamics using a coarse-grained protein model~\cite{kuriata2018cabs}. 

{\bf Protein dataset.}
We present here more details on the main set of proteins we used in this work.

{\it HRas.}
HRas plays an important role in signal transduction during cell-cycle regulation~\cite{mccormick1995ras}. Previous studies have shown that calcium acetate acts as an allosteric activator and its mechanism of allostery is mediated by a network of hydrogen bonds, involving structural water molecules, that link the allosteric site to the catalytic residue Q61~\cite{buhrman2010allosteric}. We treat the allosteric and active sites, that are located at opposite ends of the protein (PDB ID: 3K8Y), as the source or target nodes in our relative dimension (since multiple atoms compose the allosteric and active sites, we use all nodes as the source of the diffusive process with a uniform distribution on them). 

{\it Lactose repressor (lac).}
As a second example, we examine the well-studied lactose repressor (lac) (PDB ID: 1EFA)~in Figure~\ref{fig:figure_allostery}b, present in E. coli and which binds to the lac operon, a section of DNA, to inhibit the expression of proteins for the metabolism of lactose when no lactose is present~\cite{becker2014bacterial,wilson2007lactose}. In its complete form, it consists of $4$ monomers, with two binding sites to a single DNA strand, inhibiting the genes located between them. The combination of two monomers co-operate to form one of the two binding sites (orange region in Figure~\ref{fig:figure_allostery}b). On each monomer there is an allosteric site for the binding of NPF molecules that activate the lac repressor.

{\it PDK1.}
PDK1 is a well-known protein Kinase (PDB ID: 3ORX) that is implicated in the progression of Melanoma's~\cite{sadowsky2011turning}. 
The allosteric site of PDK1 is a sequence of amino acids, called the PDK1 interacting fragment (PIF), that binds to a phosphate on the catalytic domain. 
This binding triggers a signal to start the phosphorylation of the activation loop of the substrates at the ATP pocket, or active site~\cite{biondi2001pif}.
The crystallographic structure (PDB ID: 3ORX) used for our analysis has the molecule BI4 bound at the active site~\cite{sadowsky2011turning} via three hydrogen bounds to a region of high relative dimension, and interacts through hydrophobic forces on a region of low relative dimension. 

{\bf Fluorescence correlation microscopy experiments.}
Protein plasmids of \emph{Aquifex} Adenylate Kinase (ID:18092 Plasmid:peT3a-AqAdk/MVGDH) were purchased from AddGene as deposited by 'Dorothee Kern Lab Plasmids'. The plasmids were already encoded with two cysteine mutations for maleimide conjugation. ADK was expressed in a 1 litre culture BL21 (DE3) cells via inoculation with 1 mM IPTG. BugBuster was used for cell lysis and TCEP and protease inhibitor was added to the lysate. ADK was purified via HIS-tag with a gravi-trap (GE-healthcare), and a PD-10 column was used to remove imidazole and exchange into protein buffer (20 mM TRIS, 5 0mM NaCl). TCEP and protease inhibitor were added throughout the purification process. Alexa 488-labelled ADK was prepared overnight using 20 $\mu$M protein with molar ratio 1:10 of protein:Alexa 488. Excess dye was removed using HIS-tag purification and a PD-10 column. A Typhoon was used to examine the gel of the purified-labelled ADK product and showed no excess fluorophore. 
The label sites for the FRET experiment were Tyr 52 (AMP$_\text{bd}$ domain) changed to Cys and Val 145 changed to Cys (lid domain) \cite{Henzler-Wildman2007a}. 
Samples were diluted to 200 pM in pH 7.5 FRET buffer (20 mM TRIS, 50mM NaCl) with 0.3 mg/ml BSA to prevent surface adsorption. Measurements were taken at thermal equilibrium such that all processes under analysis are statistical fluctuations around the equilibrium.
Freely diffusing single-molecules were detected using a home-built dual-channel confocal fluorescence microscope. A tunable wavelength argon ion laser (model 35LAP321-230, Melles Griot, Carlsbad, CA) was set to 514.5 nm to excite Alexa 488. The beam was focused into the sample solution to a diffraction-limited spot with a high numerical aperture oil-immersion objective (Nikon Plan Apo TIRF 60x, NA 1.45). The closer refractive indexes of oil and glass relative to water and glass make oil immersion preferable due to reduced light reflection. Type FF immersion oil (Cargille, USA) was used due to its negligible fluorescent properties.
The obtained fluctuations of fluorescence intensity are autocorrelated. 
We fit the autocorrelation curves with a global model that includes components for triplet excitation, conformational dynamics and diffusion, with the assumption that they differed by a factor of 1.6 to distinguish the components,
\begin{align*}
G(\tau) = G(0)\left (\frac{1}{1+\frac{\tau}{\tau_{D}}}\right ) \left ( 1 - F + Fe^{\frac{\tau}{\tau_{m}}} \right ) \\ \left ( 1 - F_2 + F_2e^{\frac{\tau}{\tau_{conf}}} \right )\, , 
\label{eq:autocorrelationfit}
\end{align*}
where $\tau_{c}$, $\tau_{m}$ and $\tau_{D}$ are the dynamical timescales of the protein conformational dynamics, mean triplet relaxation and the protein diffusion respectively. $F_{1}$ is the fraction of molecules entering the triplet state and $F_{2}$ is the fraction of molecules conformationally fluctuating.

{\bf Root-mean square fluctuation analysis.} 
We use the cabs flex 2 server that generated fast simulations of near-native dynamics. The dynamics uses Monte Carlo dynamics and an asymmetric metropolis scheme. CABS is a well established coarse grained (i.e. atoms are combined into larger units) protein modelling tool.
CABS uses a forcefield derived from statistical regularities seen in known protein structures, and it includes side-chain-side-chain mean field potentials, coarse-grained models of main chain hydrogen bonds, and local peptide-chain geometric preferences. The solvent effect is accounted for in an implicit fashion through protein structure statistics used in the derivation of the CABS force-field. The dynamics of CABS-based coarse-grained proteins is simulated by a random series of local conformational transitions (controlled by a Monte Carlo method).  The results show strong similarities with fully atomistic MD simulations.
(Description here \url{http://biocomp.chem.uw.edu.pl/sites/default/files/publications/ct300854w.pdf} )
The resulting trajectory from the MD simulation is analysed and clustered to a representative ensemble of protein models that reflect the flexibility of the input structure.
In short, the simulation (like other MD simulations) examines the dynamic evolution of interacting units (atoms or coarse grained units). The trajectories are determined by solving Newtons equations of motion, where the forces between units are determined by the proposed forcefield. Therefore, inherently one can study the thermodynamic properties of a system via a MD simulation.

{\bf SIR model.}
For the example with SIR dynamics, we simulated the standard SIR model on networks, using
the fast approximation of~\cite{kiss2017mathematics}, with open sourced code available at \url{https://github.com/springer-math/Mathematics-of-Epidemics-on-Networks} and estimated the infectiousness of each node as the averaged number of removed nodes when the spread started from this node over $500$ realisation of the dynamics. 
To estimate the critical value for the infectiousness $\beta$, we computed the average infectability across all nodes for each $\beta$ and estimated  $\beta_{\rm crit}$ as the value for which half of the nodes are infected.

{\bf Graph classification dataset.}
We generated $600$ graphs of each of the three classes, Erdos-Renyi, Barabasi-Albert and Small Worlds.
We sampled the number of nodes with $10$ bins from $100$ to $1000$, and repeated that $3$ times with different random seed. For in each case, we created $20$ networks of each types with the following range of parameters: ER from with probabilities from $0.03$ to $0.1$, BA with number of edges per nodes from $1$ to $20$ and SW with probability from $0.1$ to $0.7$ and number of neighbours from $5$ to $10$. Improvements to the random-graph classification results can be made using other graph theoretic features~\cite{peach2021hcga}.

\section{Code Availability}

The code is shared under the GNU General Public License v3.0. It can be found at \url{https://github.com/barahona-research-group/DynGDim} and \url{https://doi.org/10.5281/zenodo.6496778}~\cite{peach_robert_2022_6496779}.

\section{Acknowledgments}

We thank David Infield, Thomas Higginson, Francesca Vianello, Florian Song, Paul Expert, Asher Mullokandov and Sophia Yaliraki for valuable discussions. 
We acknowledge funding through EPSRC award EP/N014529/1 supporting the EPSRC Centre for Mathematics of Precision Healthcare at Imperial. RP acknowledges funding from the Deutsche Forschungsgemeinschaft (DFG, German Research Foundation) Project-ID 424778381-TRR 295.
AA was supported by funding to the Blue Brain Project, a research center of the École polytechnique fédérale de Lausanne (EPFL), from the Swiss government’s ETH Board of the Swiss Federal Institutes of Technology.

\bibliography{transiplus}
\widetext

\section{Supplementary Information}
\section{Degree distributions}\label{supp:degree}

\begin{figure*}
    \centering
    \includegraphics[width=1\textwidth]{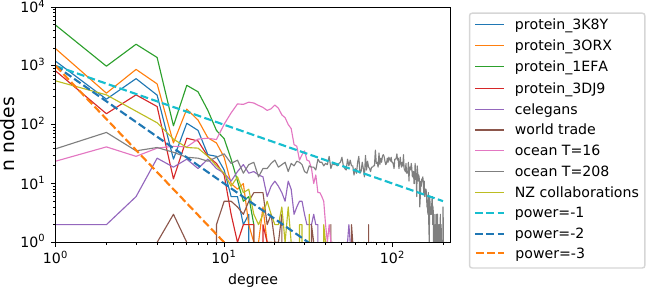}
    \caption{\textbf{Supplementary Figure 1} Degree distributions of some of the networks considered in this manuscript. Isolines corresponding to power-law distributions are shown as dashed lines.}
    \label{fig:SI_degree}
\end{figure*}

We display the degree distributions of the main networks used within this manuscript in Figure~\ref{fig:SI_degree}. Our measure of dimension is applicable to networks of any structure, including those with non-trivial topological features, namely complex networks. We find that the protein networks display a distribution that is resembles a power-law, and thus can be considered to have some `scale-free' properties. Other networks can be considered more homogeneous in terms of their degree distribution, including the NZ collaboration social network.

\section{C. Elegans}

Analysis of the undirected connectome (N=377) of the nematode \textit{C.\ elegans} with the inclusion of muscles, important for examining control~\cite{yan2017network}\footnote{\url{https://www.wormatlas.org/neuronalwiring.html}}, and where scales have previously been shown as important~\cite{bacik2016flow}. We highlight the top $40$ local dimension neurons in Table~\ref{tbl:neurons}.

\begin{table}[]
\begin{tabular}{llll}
\hline
Neuron & Neuron type & local\_dimension   & z-score            \\ \hline
AVAL   & I           & 2.563 & 2.372 \\
AVBL   & I           & 2.554 & 2.304 \\
AVAR   & I           & 2.545 & 2.238  \\
AVBR   & I           & 2.543 & 2.224 \\
AVER   & I           & 2.536 & 2.171 \\
RICL   & I           & 2.520 & 2.050 \\
ADAR   & I           & 2.513 & 1.998 \\
AVDL   & I           & 2.512 & 1.988 \\
RIGL   & I           & 2.512 & 1.987 \\
AVEL   & I           & 2.509 & 1.964 \\
AVDR   & I           & 2.508 & 1.956 \\
RIS    & I           & 2.504 & 1.929  \\
RMGL   & M           & 2.494 & 1.850 \\
RICR   & I           & 2.488 & 1.807 \\
ADEL   & S           & 2.485 & 1.787 \\
RMGR   & M           & 2.482 & 1.759 \\
RIML   & M           & 2.479 & 1.739  \\
SDQR   & I           & 2.479 & 1.739 \\
RIMR   & M           & 2.475 & 1.712 \\
ADER   & S           & 2.475 & 1.712 \\
PVCL   & I           & 2.474 & 1.704 \\
AVKL   & I           & 2.470 & 1.670 \\
PVCR   & I           & 2.470 & 1.669  \\
ALA    & I           & 2.467 & 1.650 \\
AVKR   & I           & 2.460 & 1.595 \\
RIGR   & I           & 2.455 & 1.561 \\
SMDDR  & M           & 2.451 & 1.528 \\
SAAVR  & I           & 2.449 & 1.513 \\
RIAR   & I           & 2.446 & 1.493 \\
ALML   & S           & 2.441 & 1.452 \\
PVNR   & I           & 2.431 & 1.374 \\
SAAVL  & I           & 2.429 & 1.365 \\
RIBR   & I           & 2.429 & 1.361 \\
RIAL   & I           & 2.428 & 1.355 \\
RIBL   & I           & 2.427 & 1.350 \\
ADAL   & I           & 2.419 & 1.286 \\
AIZR   & I           & 2.416 & 1.267 \\
FLPR   & S           & 2.415 & 1.259 \\
PVPR   & I           & 2.415 & 1.256 \\
SABD   & I           & 2.413 & 1.243 \\ \hline
\end{tabular}
\caption{\textbf{Supplementary Table 1} The 40 neurons with the highest local dimension. The z-score is calculated in respect to the mean and standard deviation of the full set of neurons.}
\label{tbl:neurons}
\end{table}

\end{document}